\documentclass[aps,pra,reprint,notitlepage,superscriptaddress,longbibliography,nofootinbib,floatfix]{revtex4-2}

\usepackage[utf8]{inputenc}
\usepackage[T1]{fontenc}
\usepackage{lmodern}

\usepackage{amsmath,amssymb,amsfonts}
\usepackage{physics}         
\usepackage{graphicx}        
\usepackage{hyperref}        
\usepackage{bm}              
\usepackage{mathtools}       
\usepackage{braket}          
\usepackage{xcolor}          
\usepackage{comment}
\usepackage{microtype}       
\usepackage{siunitx}         
\usepackage{enumitem}        

\usepackage{tikz}
\usepackage{amsmath,amssymb}
\usetikzlibrary{decorations.pathmorphing,decorations.pathreplacing}
\usepackage{pgfplots}
\usetikzlibrary{arrows.meta, decorations.markings}

\begin{document}

\title{Entropy Flow and Exceptional-Point Structure in Two-Mode Squeezed-Bath Dynamics
}

\author{Eric R Bittner}
\affiliation{Department of Physics, University of Houston, Houston, TX 77204, USA}



\date{\today}

\begin{abstract}
Squeezed reservoirs provide a powerful means of engineering nonclassical noise and controlling irreversible dynamics in open quantum systems. Here we develop a comprehensive analysis of two coupled harmonic oscillators driven by independent squeezed baths, focusing on the emergence of coherence-driven entropy flow and the structure of exceptional points (EPs) in the corresponding Lindblad dynamics. Working entirely within the Gaussian formalism, we derive closed-form evolution equations for the covariance matrix and show that squeezing induces entropy generation only at \emph{second order} in the anomalous correlations—a nonlinear mechanism absent in thermal environments. This entropy flow is accompanied by a rich non-Hermitian structure: by scanning the squeezing parameters we uncover a characteristic “exceptional-point fan’’ in the $(M_{1}, M_{2})$ plane, which separates a narrow PT-unbroken region of oscillatory dynamics from broad PT-broken sectors in which one normal mode becomes purely overdamped. This geometric organization of EPs reveals that PT symmetry survives only when the two reservoirs squeeze opposite quadratures, and is generically broken for in-phase squeezing. Our analysis establishes squeezed reservoirs as a natural setting where information-bearing noise drives irreversible behavior through coherent pathways, and lays the groundwork for experimentally accessible probes of entropy flow and critical mode behavior in more complex open systems.
\end{abstract}

\maketitle

\section{Introduction}

The study of open quantum systems traditionally models environments as collections of harmonic oscillators in thermal equilibrium. Recent advances in quantum optics and reservoir engineering have made it possible to prepare systems coupled to \emph{squeezed baths}—environments that carry intrinsic quantum coherences between modes. These baths suppress fluctuations along one quadrature while amplifying fluctuations along the conjugate one, a property that strongly modifies system dynamics and steady states.

Squeezed thermal reservoirs arise naturally in parametrically driven cavities, non-degenerate parametric amplifiers, and engineered microwave environments in superconducting circuits. They enable novel dissipative regimes in which quantum coherence and entanglement can persist, emerge, or even be enhanced by noise. Prior work has explored squeezed environments in quantum optics, cavity QED, and nonequilibrium thermodynamics. Gardiner and Collett's quantum input-output theory \cite{gardiner1985input} established how squeezed vacuum fields modify system observables, while Caves \cite{caves1981quantum} showed that squeezed light can surpass standard quantum limits in precision measurement. Kronwald, Marquardt, and Clerk \cite{kronwald2013arbitrarily} demonstrated that engineered reservoirs can steer systems toward highly nonclassical steady states. More recently, Metelmann and Clerk \cite{metelmann2015nonreciprocal} showed that squeezed dissipation can break reciprocity and enable directional amplification. Other studies have extended squeezed-bath thermodynamics to quantum heat engines, revealing that squeezing serves as a source of free energy that alters entropy production in nontrivial ways \cite{Arisoy2022}. Collectively, these works underscore the fundamental connections between squeezing, irreversibility, and information flow in open quantum systems.

Experimentally, squeezed reservoirs can be realized by coupling a system to the output of a degenerate or non-degenerate parametric amplifier. In optical platforms, this is typically achieved by operating an optical parametric oscillator below threshold, where a nonlinear medium produces squeezed vacuum states. In superconducting circuit QED, Josephson parametric amplifiers and traveling-wave parametric amplifiers generate squeezed microwave fields \cite{mallet2011quantum,flurin2012generating}. These fields can be routed through circulators and injected into a cavity or waveguide coupled to a qubit or resonator, thereby realizing a controlled squeezed reservoir.

\begin{figure}[htbp]
    \centering
    \includegraphics[width=\columnwidth]{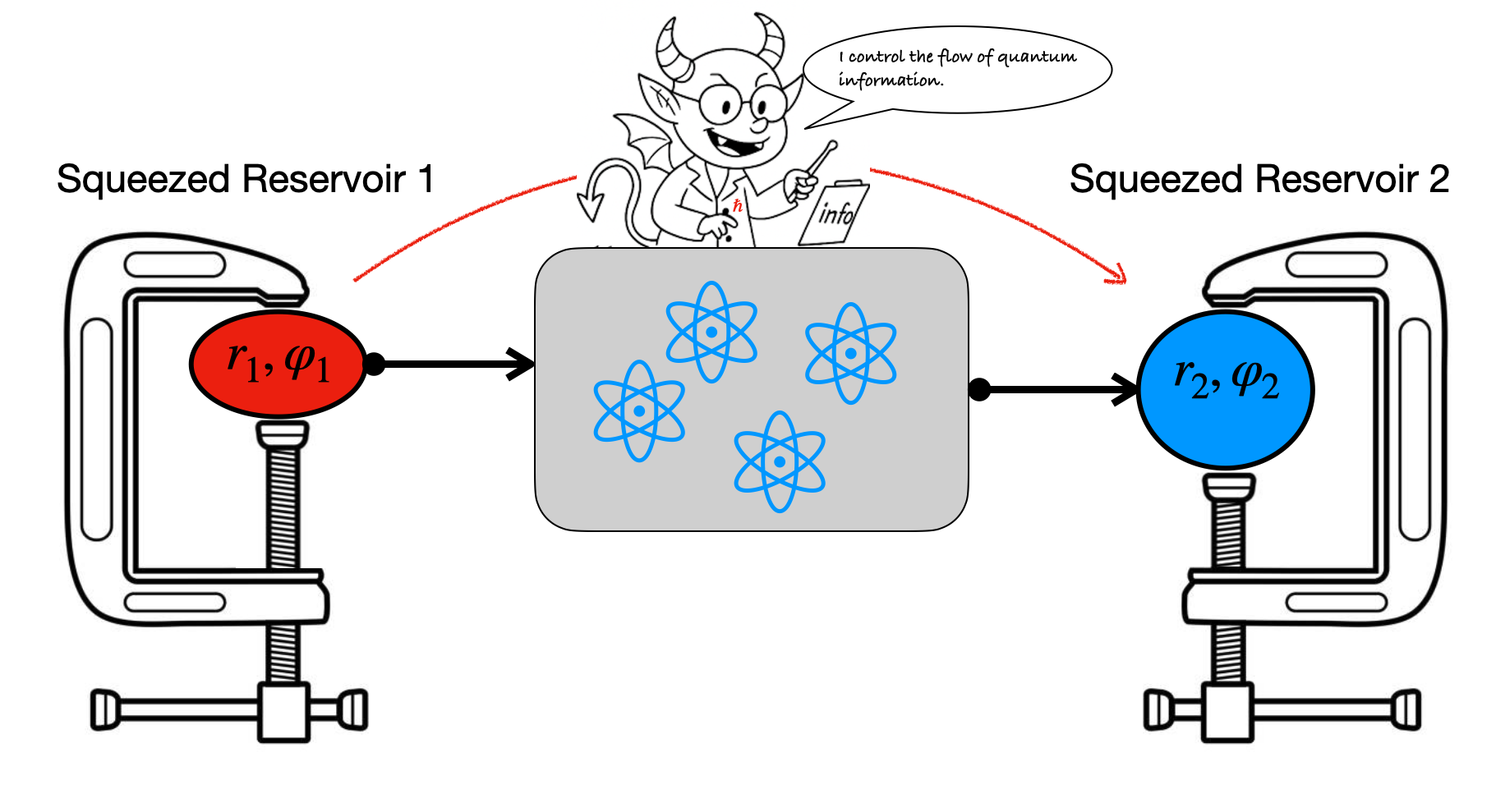}
    \caption{
    \textbf{Conceptual illustration of a quantum system coupled to two local squeezed reservoirs. } Each reservoir acts as a phase-sensitive environment characterized by squeezing parameters \( (r_1, \varphi_1) \) and \( (r_2, \varphi_2) \), shown here as mechanical clamps exerting quadrature-specific noise. The central system consists of interacting bosonic modes. Asymmetry in the squeezing profiles leads to a directional flow of entropy and coherence, even in the absence of a thermal gradient. The cartoon demon highlights the role of information in structuring quantum noise and emphasizes the analogy to feedback-driven thermodynamic processes.
    }
    \label{fig:squeezed_demon}
\end{figure}

The purpose of this work is to develop a compact and analytic framework describing how squeezed reservoirs influence the dynamics of linear and weakly nonlinear quantum systems. We begin with a single bosonic mode coupled to a squeezed bath and extend our analysis to multimode systems where squeezed dissipation competes with coherent tunneling and detuning. Our goal is to clarify how squeezed noise reshapes relaxation pathways, alters asymptotic purity, and introduces spectral features—such as exceptional points—that signify transitions in dynamical behavior.

Although squeezed reservoirs have been extensively studied in contexts such as quantum control, enhanced cooling, and work extraction \cite{Gardiner2004,Zippilli2014,Murch2013,Klaers2017,Alicki2015}, their role in shaping the spectral structure of dissipative dynamics remains comparatively underexplored. We show that the drift matrix governing the first and second moments of two coupled oscillators develops exceptional points (EPs) as a function of the squeezing parameters and intermode coupling. These EPs—non-Hermitian degeneracies where eigenvalues and eigenvectors coalesce—are identified by the divergence of the condition number of the eigenvector matrix \cite{Heiss2012,Minganti2019}. Our results reveal that squeezing can drive the system across these spectral transitions, producing distinct dynamical regimes characterized by strong mode mixing and enhanced sensitivity.

To our knowledge, this work provides the first demonstration of exceptional-point structure induced by squeezed-reservoir dissipation within the Lindblad/Markov open-quantum-system formalism. It establishes a direct link between squeezing-driven spectral transitions, coherence loss, and entropy generation. This contrasts with previous studies that examined exceptional points arising from parametric driving or unsqueezed dissipative processes.


\section{Theoretical Framework}

\subsection{System–Bath Hamiltonian and Squeezed Bath Properties}

We consider a quantum system with degrees of freedom represented by bosonic operators \( \hat{A} \) and \( \hat{A}^\dagger \), linearly coupled to an environment composed of independent harmonic oscillators. The total Hamiltonian is
\begin{equation}
H = H_S(\hat{A}^\dagger, \hat{A}) + \sum_k \omega_k b_k^\dagger b_k + \sum_k g_k \left( \hat{A} b_k^\dagger + \hat{A}^\dagger b_k \right),
\end{equation}
where \( H_S \) is the intrinsic system Hamiltonian, \( b_k \) and \( b_k^\dagger \) are the annihilation and creation operators of bath mode \( k \) with frequency \( \omega_k \), and \( g_k \) is the coupling constant between the system and bath mode \( k \).

Although thermal baths are ubiquitous in quantum optics and statistical physics, squeezed environments arise less frequently and are often unfamiliar to a broader audience. Because they play a central role in our analysis, we now review their structure, origin, and characteristic properties. A squeezed bath introduces nonclassical correlations among its constituent modes, alters the fluctuation–dissipation balance, and modifies the noise spectrum seen by the system. These features profoundly impact the system’s steady-state behavior, coherence properties, and response to external driving.
The usual thermal density matrix of a bosonic mode \( b_k \) with average occupation number \( n_k \) can be written in exponential form as
\begin{equation}
\rho_{\mathrm{th}}^{(k)} = \frac{1}{Z_k} e^{-\beta \omega_k b_k^\dagger b_k}, \qquad n_k = \frac{1}{e^{\beta \omega_k} - 1},
\end{equation}
with partition function \( Z_k = \mathrm{Tr}(e^{-\beta \omega_k b_k^\dagger b_k}) = \frac{1}{1 - e^{-\beta \omega_k}} \).
Equivalently, one can express this state in the following operator-normal-ordered form:
\begin{equation}
\rho_{\mathrm{th}}^{(k)}(n_k) = \frac{1}{n_k + 1} \left( \frac{n_k}{n_k + 1} \right)^{b_k^\dagger b_k}.
\end{equation}

Unlike conventional thermal reservoirs, we assume that each bath mode is initialized in a squeezed thermal state. That is, rather than the standard thermal density matrix, each mode \( b_k \) is in the state
\begin{equation}
\rho_B^{(k)} = S_k(r_k, \phi_k) \, \rho_{\text{th}}^{(k)} \, S_k^\dagger(r_k, \phi_k),
\end{equation}
where \( S_k(r_k, \phi_k) \) is the single-mode squeezing operator,
\begin{equation}
S_k(r_k, \phi_k) = \exp\left[ \frac{1}{2} r_k \left( e^{-2i\phi_k} b_k^2 - e^{2i\phi_k} b_k^{\dagger 2} \right) \right].
\end{equation}
Here, \( r_k \geq 0 \) is the squeezing amplitude that determines the degree of noise suppression, and \( \phi_k \in [0, 2\pi) \) is the squeezing phase that orients the noise ellipse in phase space.
The squeezing operator also modifies the thermal state by introducing nonzero anomalous correlations. The resulting squeezed thermal state is a Gaussian state completely characterized by its second-order moments. In particular, it possesses a modified mean occupation number
\begin{equation}
N_k = \langle b_k^\dagger b_k \rangle = \bar{n}_k \cosh(2r_k) + \sinh^2 r_k,
\label{eq:6}
\end{equation}
as well as an anomalous correlation
\begin{equation}
M_k = \langle b_k b_k \rangle  =-\frac{1}{2} \sinh(2r_k) e^{2i\phi_k}(2\bar{n}_k + 1),
\label{eq:7}
\end{equation}
 These two quantities completely specify the Gaussian statistics of the reservoir in the Markovian limit and will serve as the primary variables throughout our analysis.
 Throughout this work, we denote by $r_k$ the squeezing amplitude, $\varphi_k$ the squeezing phase, and $M_k = |M_k|e^{i\varphi_k}$ the complex anomalous correlation parameter that fully characterizes the squeezing strength. Unless stated otherwise, $M$ refers to the effective complex squeezing parameter of a given bath.
Similarly, where appropriate, we will refer to \( N \) as the effective thermal occupation number and to \( M \) as the anomalous coherence parameter, with the physical constraint \( |M|^2 \leq N(N+1) \) ensuring positivity of the bath's density matrix and compliance with the uncertainty principle.  Going forward, we shall use \( N_k \) and \( M_k \) to parameterize the influence of the bath on the system dynamics. These two quantities will enter explicitly in the master equation and determine the modified fluctuation–dissipation relation for systems coupled to squeezed reservoirs

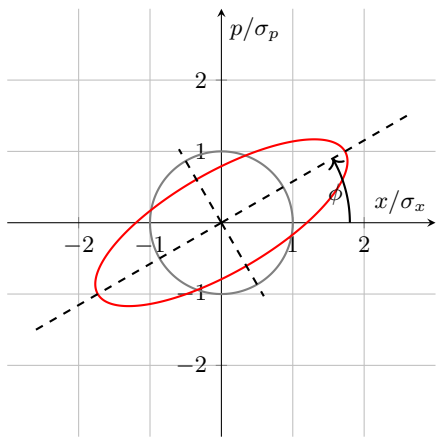
\begin{figure}[b]
    \centering
\begin{tikzpicture}
  \begin{axis}[
      axis equal image,
      xlabel={$x/\sigma_x$},
      ylabel={$p/\sigma_p$},
      xmin=-3, xmax=3,
      ymin=-3, ymax=3,
      axis lines=middle,
      xtick={-2,-1,0,1,2},
      ytick={-2,-1,0,1,2},
      grid=both,
      clip=false  
    ]

    \def\a{2}  
    \def\b{0.7}  
    \def\ph{30} 

    \addplot+[domain=0:2*pi, samples=200, thick, gray, mark=none]
      ({cos(deg(x))}, {sin(deg(x))});

    \addplot+[domain=0:2*pi, samples=200, thick, red, mark=none]
      ({\a*cos(deg(x))*cos(\ph) - \b*sin(deg(x))*sin(\ph)},
       {\a*cos(deg(x))*sin(\ph) + \b*sin(deg(x))*cos(\ph)});

    \addplot[dashed, thick, black]
      coordinates {(0,0) 
        ({1.5*\a*cos(\ph)},{1.5*\a*sin(\ph)})};

    \addplot[dashed, thick, black]
      coordinates {(0,0) 
        ({-1.5*\a*cos(\ph)},{-1.5*\a*sin(\ph)})};
    \addplot[dashed, thick, black]
      coordinates {(0,0) 
        ({-1.7*\b*sin(\ph)},{1.7*\b*cos(\ph)})};
    \addplot[dashed, thick, black]
      coordinates {(0,0) 
        ({1.7*\b*sin(\ph)},{-1.7*\b*cos(\ph)})};

    \addplot[domain=0:30, samples=40, thick, ->]
      ({1.8*cos(x)}, {1.8*sin(x)});

    \node at (axis cs:1.6,0.4) {$\phi$};
  \end{axis}
\end{tikzpicture}
\caption{
Wigner representation of thermal (gray circle) and squeezed (red ellipse) states for a single-mode bosonic system. 
The horizontal and vertical axes correspond to the dimensionless position and momentum quadratures 
\(x\) and \(p\), respectively. 
The unit lengths along each axis are set by the thermal variances 
\(\sigma_x = \sqrt{\langle \hat{x}^2 \rangle}\) and 
\(\sigma_p = \sqrt{\langle \hat{p}^2 \rangle}\), 
such that the thermal (unsqueezed) state appears as a unit circle. 
Squeezing distorts this distribution into an ellipse with principal axes oriented at angle \(\phi\), corresponding to the squeezing phase. 
The arc marks this squeezing angle relative to the \(x\)-axis.
}
    \label{fig:wigner-squeezing}
\end{figure}

%
%
Further, they define the bath’s statistical and quantum fluctuations through the following time-domain correlation functions:
\begin{align}
\ev{b_k^\dagger(t) b_{k'}(t')} &= \delta_{kk'} N_k e^{i\omega_k(t - t')}, \\
\ev{b_k(t) b_{k'}^\dagger(t')} &= \delta_{kk'} (N_k + 1) e^{-i\omega_k(t - t')}, \\
\ev{b_k(t) b_{k'}(t')} &= \delta_{kk'} M_k e^{-i\omega_k(t + t')}, \\
\ev{b_k^\dagger(t) b_{k'}^\dagger(t')} &= \delta_{kk'} M_k^* e^{i\omega_k(t + t')}.
\end{align}
The first two expressions reproduce standard thermal noise properties, while the last two impose anomalous correlations that reflect coherent interactions between pairs of creation or annihilation events.

\begin{widetext}

Equivalently, the squeezed thermal state for each bath mode \( k \) can be described by its Wigner function in phase space:
\begin{equation}
W_k(x, p) = \frac{1}{2\pi \sqrt{\det \boldsymbol{\sigma}_k}} \exp\left( -\frac{1}{2} 
\begin{bmatrix}
x & p
\end{bmatrix}
\boldsymbol{\sigma}_k^{-1}
\begin{bmatrix}
x \\ p
\end{bmatrix}
\right),
\end{equation}
where \( \boldsymbol{\sigma}_k \) is the covariance matrix of the squeezed thermal state. For a squeezing parameter \( r_k \), angle \( \phi_k \), and thermal occupation \( n_k \), the covariance matrix takes the form
\begin{align*}
\boldsymbol{\sigma}_k = \left(n_k + \tfrac{1}{2}\right)
\begin{bmatrix}
\cosh(2r_k) - \sinh(2r_k)\cos(2\phi_k) & -\sinh(2r_k)\sin(2\phi_k) \\
-\sinh(2r_k)\sin(2\phi_k) & \cosh(2r_k) + \sinh(2r_k)\cos(2\phi_k).
\end{bmatrix}
\end{align*}

This clearly illustrates that squeezing modifies both the magnitude and orientation of fluctuations in the \( x \)–\( p \) plane, with minimal uncertainty ellipses rotated by the squeezing angle \( \phi_k \) and stretched or compressed according to \( r_k \).
To provide a geometric intuition for how squeezing modifies the state of the bath, we illustrate in Fig.~\ref{fig:wigner-squeezing} the Wigner distributions of a thermal state (gray circle) and a squeezed state (red ellipse) in phase space. The axes correspond to the dimensionless position and momentum quadratures \( x \) and \( p \), with unit lengths set by their respective thermal variances, \( \sigma_x = \sqrt{\langle \hat{x}^2 \rangle} \) and \( \sigma_p = \sqrt{\langle \hat{p}^2 \rangle} \), such that an unsqueezed (thermal) state appears as a circle. The squeezed state is represented by an ellipse whose principal axes reflect reduced noise along one quadrature and enhanced noise along the orthogonal direction. The orientation of this ellipse is determined by the squeezing phase \( \phi \), marked in the figure by an arc. While thermal states are isotropic and solely governed by temperature (encoded by the parameter \( N \)), squeezed states exhibit anisotropic correlations encoded by both the squeezing magnitude \( r \) and angle \( \phi \).

\end{widetext}

\subsection{Lindblad Equations}
To derive the effective dynamics of the system alone, we adopt the standard Born--Markov approximation. This involves two key assumptions that the system–bath coupling is weak enough that the joint state can be factorized at all times as $\rho_{\text{tot}}(t) \approx \rho(t) \otimes \rho_B$, where $\rho(t)$ is the system's reduced density matrix and $\rho_B$ is the stationary bath state, and that the bath correlation time is much shorter than any intrinsic time-scale of the system allowing us to neglect memory effects and extend integrals over the past history of the system to $t \to \infty$ (Markov approximation).  
In recent analyses, it has been shown that non-Markovian effects in squeezed reservoirs lead to modified relaxation dynamics 
and information backflow, altering coherence lifetimes \cite{Ablimit2023}. 
Such effects may contribute to the transient entropy production discussed in the present model.
Under these assumptions, the second-order perturbative expansion of the interaction-picture evolution yields the general form:
\begin{equation}
\frac{d}{dt} \rho(t) = -\int_0^\infty d\tau \, \text{Tr}_B \left[ H_{\text{int}}(t), \left[ H_{\text{int}}(t - \tau), \rho(t) \otimes \rho_B \right] \right],
\end{equation}
where all operators are expressed in the interaction picture with respect to $H_0 = H_S + H_B$.

Substituting the system–bath coupling $H_{\text{int}} = \sum_k g_k (A b_k^\dagger + A^\dagger b_k)$ and inserting the squeezed-bath correlation functions, one obtains additional terms in the master equation beyond the standard thermal Lindblad form. Specifically, the anomalous correlations $\ev{b_k(t) b_k(t')}$ and $\ev{b_k^\dagger(t) b_k^\dagger(t')}$ introduce squeezing-induced dissipators of the form
\begin{align}
\mathcal{D}_M[\rho] &= \gamma M \left(A \rho A - \tfrac{1}{2}\{A^2, \rho\}\right)\nonumber \\
&+ \gamma M^* \left(A^\dagger \rho A^\dagger - \tfrac{1}{2}\{(A^\dagger)^2, \rho\}\right),
\end{align}
which couple to the square of the system operators. These terms lead to phase-sensitive dissipation and can give rise to coherent steady states or reduced entropy production.

We emphasize that the full master equation retains the standard thermal dissipators:
\begin{equation}
\mathcal{L}_{\text{th}}[\rho] = \gamma (N + 1) \mathcal{D}[A]\rho + \gamma N \mathcal{D}[A^\dagger]\rho,
\end{equation}
so that the total Liouvillian becomes
\begin{equation}
\frac{d\rho}{dt} = -i[H_S, \rho] + \mathcal{L}_{\text{th}}[\rho] + \mathcal{D}_M[\rho].
\end{equation}

The additional dissipators $\mathcal{D}_M[\rho]$ introduced by squeezing are fundamentally distinct from thermal noise in that they break the phase symmetry of the bath. These terms correspond to processes where the bath mediates pairwise creation or annihilation of system excitations. As such, they favor states that are squeezed along quadratures determined by the phase of $M$, and they can either purify or destabilize the steady state, depending on the interplay between $M$, $N$, and the system Hamiltonian.

Despite breaking number conservation, the dynamics remain completely positive as long as the bath parameters satisfy:
\begin{equation}
|M|^2 \leq N(N + 1),
\end{equation}
which ensures the underlying Gaussian state is physical.

\begin{figure}[htbp]
  \centering
  \includegraphics[width=\columnwidth]{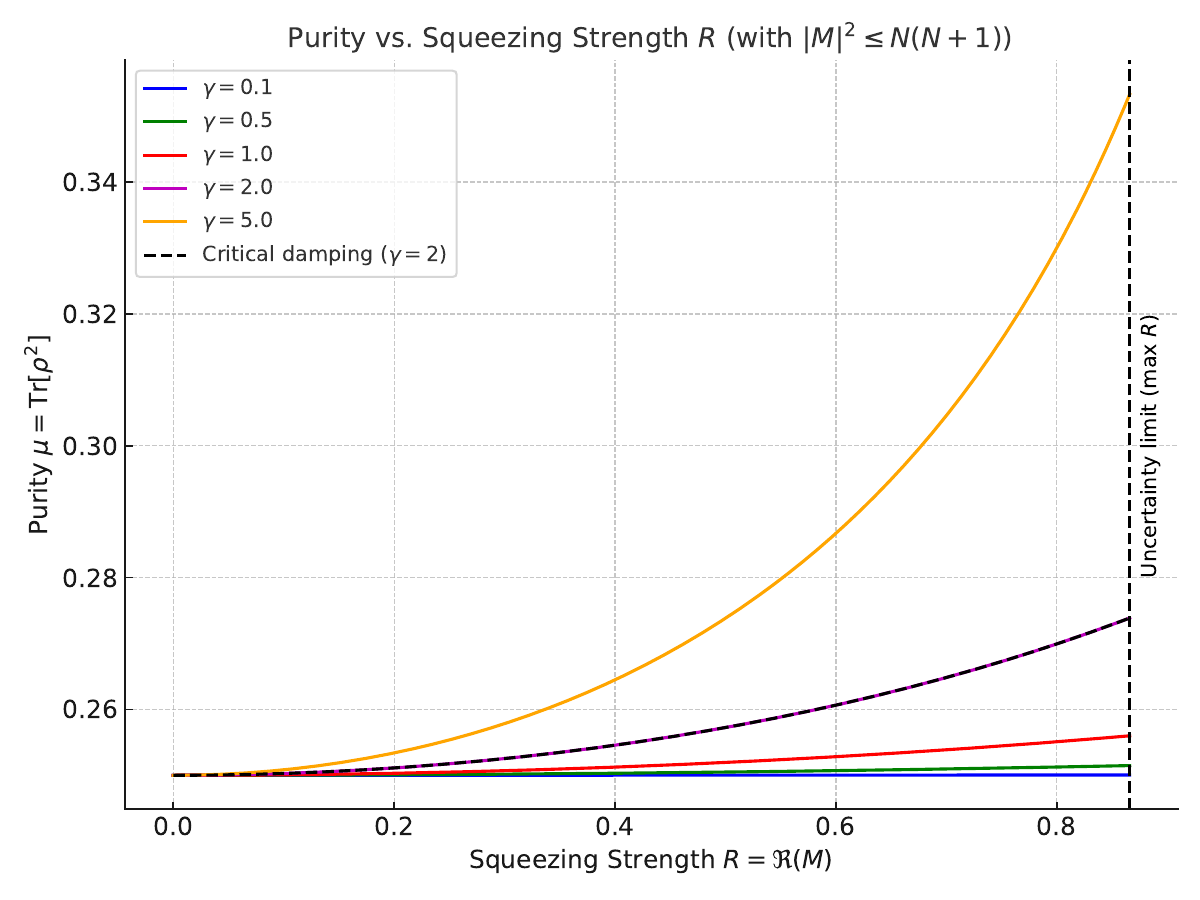}
  \caption{
    Steady-state purity $\mu = \mathrm{Tr}[\rho^2]$ versus squeezing strength $R = \Re(M)$ for a single harmonic oscillator coupled to a squeezed thermal bath, with $N = 0.5$ and $\omega = 1$. 
    Each curve corresponds to a different value of the damping rate $\gamma$.
    The vertical dashed line marks the uncertainty-limited maximum value of $R = \sqrt{N(N+1)} \approx 0.866$, beyond which the bath state becomes unphysical.
    As $R$ increases, the system is driven to a lower-entropy, more coherent steady state.
    As the squeezing strength increases, the oscillator approaches a minimum-uncertainty state, demonstrating how phase-sensitive dissipation can purify steady-state coherence.
  }
  \label{fig:purity-vs-R}
\end{figure}

\subsection{Heisenberg Evolution and Exceptional Point Structure}

To gain insight into the dynamical symmetries of the system, we examine the Heisenberg equations of motion for the field operators $a$ and $a^\dagger$ under the influence of the squeezed bath. These are most compactly written in vector form as:
\begin{equation}
\frac{d}{dt} 
\begin{pmatrix}
a \\
a^\dagger
\end{pmatrix}
= A 
\begin{pmatrix}
a \\
a^\dagger
\end{pmatrix},
\end{equation}
where the drift matrix $A$ arises from both the Hamiltonian and dissipative contributions. Under the standard Born–Markov and secular approximations, the master equation for the reduced density operator yields the following non-Hermitian drift matrix in the operator basis:
\begin{equation}
A = 
\begin{pmatrix}
-i\omega - \frac{\gamma}{2} & -\gamma M \\
-\gamma M^* & +i\omega - \frac{\gamma}{2}
\end{pmatrix}.
\label{eq:A-matrix}
\end{equation}
Here, $\omega$ is the oscillator frequency, $\gamma$ the system–bath coupling rate, and $M$ the (complex) anomalous squeezing parameter. The off-diagonal terms proportional to $M$ reflect the presence of anomalous processes (e.g., $a a$ and $a^\dagger a^\dagger$), which are absent in a purely thermal bath. 

The eigenvalues of the drift matrix $A$ are
\begin{equation}
\lambda_{\pm} = -\frac{\gamma}{2} \pm \sqrt{ \gamma^2 |M|^2-\omega^2 },
\end{equation}
indicate that an exceptional point (EP) occurs at the critical value $|M| = M_c = \omega / \gamma$. At this point, the eigenvalues coalesce and the drift matrix becomes defective, signaling a non-diagonalizable structure. This bifurcation separates underdamped dynamics ($|M| < M_c$) from overdamped decay ($|M| > M_c$), and marks a squeezing-induced symmetry-breaking transition in the system's relaxation dynamics.
The presence of squeezing in the bath—induces mixing between $a$ and $a^\dagger$ and breaks time-reversal symmetry.

Physically, crossing this exceptional point corresponds to a transition in the system’s dissipative response. Below the threshold, the oscillator exhibits oscillatory relaxation with phase-coherent damping. Beyond it , the dynamics become overdamped and time-reversal symmetry is effectively broken. Thus, the exceptional point marks the onset of a squeezing-induced dynamical symmetry breaking between energy and coherence flow.
The connection between exceptional points in non-Hermitian Hamiltonians and those arising in open quantum Liouvillians
has been explored extensively by Minganti \textit{et al.}~\cite{Minganti2019}, who showed that quantum jumps and
dissipative processes can give rise to analogous coalescence phenomena in the Liouvillian spectrum.
Physically, the EP signifies a spontaneous breaking of dynamical symmetry and a transition in the oscillator’s dissipative response from underdamped to overdamped relaxation. This occurs at a critical squeezing strength $|M| = \omega/\gamma$, beyond which the drift matrix becomes non-diagonalizable. Importantly, this dynamical threshold is distinct from the Heisenberg uncertainty limit, which places a stricter bound $|M|^2 \leq N(N+1)$ on the allowed squeezing. If the EP lies outside this allowed range, the system may never reach it physically, as the required bath correlations become nonphysical.
The analysis of this EP structure in the single-oscillator case sets the stage for understanding similar transitions in multi-mode systems, where EPs can arise due to mode-coupling and reservoir-induced coherence, even in the absence of direct Hamiltonian interaction.

We now compute the steady-state variances of a single harmonic oscillator linearly coupled to a squeezed thermal bath with average occupation $N$ and squeezing amplitude $M$.
Using the adjoint form of the master equation, we derive the Heisenberg equations of motion for the second moments:
\begin{align}
\frac{d}{dt}\langle a^\dagger a \rangle &= -\gamma \langle a^\dagger a \rangle + \gamma N \\
\frac{d}{dt}\langle a^2 \rangle &= (-2i\omega - \gamma)\langle a^2 \rangle + \gamma M \\
\frac{d}{dt}\langle a^{\dagger 2} \rangle &= (2i\omega - \gamma)\langle a^{\dagger 2} \rangle + \gamma M^*
\end{align}
Solving in the steady-state limit ($\dot{\rho} = 0$) gives:
\begin{align}
\langle a^\dagger a \rangle &= N \\
\langle a^2 \rangle &= \frac{\gamma M}{\gamma + 2i\omega} \\
\langle a^{\dagger 2} \rangle &= \frac{\gamma M^*}{\gamma - 2i\omega}
\end{align}

Hence, the steady-state covariance matrix in the $(a^\dagger, a)$ basis is:
\begin{equation}
\sigma = \begin{pmatrix}
N + \tfrac{1}{2} & \dfrac{\gamma M}{\gamma + 2i\omega} \\
\dfrac{\gamma M^*}{\gamma - 2i\omega} & N + \tfrac{1}{2}
\end{pmatrix}
\end{equation}

This captures both thermal and squeezed contributions to the noise spectrum and forms the foundation for quantifying purity, entanglement, and mutual information in multi-mode extensions.

\subsection{Dynamics of Quadratures and Variances}

We now consider the simplest nontrivial case: a single harmonic oscillator coupled to a squeezed thermal bath. The system Hamiltonian is taken to be $H_S = \omega a^\dagger a$, where $a$ is the annihilation operator of the mode, and we apply the general master equation derived above with $S = a$ and define the canonical quadratures:
\begin{equation}
\hat{x} = \frac{1}{\sqrt{2}} (a + a^\dagger), \quad \hat{p} = \frac{1}{\sqrt{2}i} (a - a^\dagger)
\end{equation}
Using the master equation derived earlier, the equations of motion for the quadrature expectation values are given 
by 
\begin{align}
\dv{\ev{\hat{x}}}{t} &= -\frac{\gamma}{2} \ev{\hat{x}} - \omega \ev{\hat{p}} \\
\dv{\ev{\hat{p}}}{t} &= -\frac{\gamma}{2} \ev{\hat{p}} + \omega \ev{\hat{x}}
\end{align}
These describe damped simple harmonic motion 
and are unaffected by either the temperature 
or squeezing of the bath. 

On the other hand, their covariances are influenced by both the temperature and squeezing. 
\begin{align}
\dv{}{t} \ev{\hat{x}^2} &= -\gamma \ev{\hat{x}^2} - 2\omega \ev{\{\hat{x}, \hat{p}\}}\nonumber \\  &+ \gamma (2N + 1 + M + M^*) \\
\dv{}{t} \ev{\hat{p}^2} &= -\gamma \ev{\hat{p}^2} + 2\omega \ev\{{\hat{x}, \hat{p}\}} \nonumber \\ &+ \gamma (2N + 1 - M - M^*) \\
\dv{}{t} \ev{\{\hat{x}, \hat{p}\}} &= -\gamma \ev{\{\hat{x}, \hat{p}\}} + 2\omega (\ev{\hat{x}^2} - \ev{\hat{p}^2})
\end{align}
To find the steady-state variances, we set all time derivatives to zero and define $R = \Re(M)$
as a convenient parameter. 
The steady-state second-order moments then given by
\begin{align}
\langle \hat{x}^2 \rangle_{\text{ss}} &= 2N + 1 + \frac{2\gamma^2 R}{\gamma^2 + 8\omega^2} \\
\langle \hat{p}^2 \rangle_{\text{ss}} &= 2N + 1 - \frac{2\gamma^2 R}{\gamma^2 + 8\omega^2} \\
\langle \{ \hat{x}, \hat{p} \} \rangle_{\text{ss}} &= \frac{8\gamma \omega R}{\gamma^2 + 8\omega^2}
\end{align}
These expressions indicate that system squeezing and coherence depend on both the damping rate and oscillator frequency. The squeezing (asymmetry in variances) vanishes in the limit $R \to 0$, and also in the overdamped ($\gamma \gg \omega$) or high-frequency ($\omega \gg \gamma$) limits.

This analysis reveals that the steady-state of the system corresponds to a squeezed thermal state. The key signature of this is the asymmetry in the quadrature variances: $\ev{\hat{x}^2} \neq \ev{\hat{p}^2}$ whenever the squeezing amplitude $M$ is nonzero. In particular, the squeezing axis is determined by the sign of the real part of $M$: when $\Re(M) > 0$, the squeezing is aligned along the $\hat{x}$ quadrature, while for $\Re(M) < 0$, the squeezing occurs along the $\hat{p}$ direction. Importantly, the state remains Gaussian at all times due to the linearity of the equations of motion and the Gaussian nature of the noise. In the special case where the squeezing saturates the uncertainty bound, i.e., $|M|^2 = N(N+1)$, the system attains a minimum-uncertainty squeezed state—analogous to an ideal optical squeezed vacuum but embedded within a thermal background.

The unequal variances in $\hat{x}$ and $\hat{p}$ not only signify the presence of squeezing in the steady state but also reflect a redistribution of quantum uncertainty between conjugate observables. While purely thermal dissipation would drive the system toward an isotropic Gaussian state with equal variances and maximal entropy for a given energy, squeezing induces coherence and directional suppression of fluctuations. These features influence not just the shape of the Wigner distribution, but also the overall purity of the state. A convenient way to quantify this is through the R\'enyi-2 entropy, which is directly related to the state’s purity via $P = \mathrm{Tr}(\rho^2) = 1/\sqrt{4 \det \sigma}$ for Gaussian states.

To compute the R\'enyi-2 entropy with these  variances, we use the general expression for a single-mode Gaussian state:
\begin{equation}
S_2 = \frac{1}{2} \ln \left[ 4\det V \right]
\end{equation}
where the $V$ is covariance matrix defined above. 
Using 
\begin{align*}
A &= 2N + 1 + \frac{2\gamma^2 R}{\gamma^2 + 8\omega^2} \\
B &= 2N + 1 - \frac{2\gamma^2 R}{\gamma^2 + 8\omega^2} \\
C &= \frac{8\gamma \omega R}{\gamma^2 + 8\omega^2}
\end{align*}
we find:
\begin{equation}
\det V = AB - \frac{1}{4} C^2 = \left(2N + 1\right)^2 - \frac{4 R^2 (\gamma^2 - 8\omega^2)^2}{(\gamma^2 + 8\omega^2)^2}
\end{equation}
Thus:
\begin{equation}
S_2 = \frac{1}{2} \ln \left[ 4\left( (2N + 1)^2 - \frac{4 R^2 (\gamma^2 - 8\omega^2)^2}{(\gamma^2 + 8\omega^2)^2} \right) \right]
\end{equation}
This entropy expression captures both quadrature squeezing and the emergence of coherence in the steady state. It also shows that maximal squeezing ($R \to N(N+1)^{1/2}$) yields minimum entropy only in the underdamped limit.

In Fig.~\ref{fig:purity-vs-R}, we show the steady-state purity of a single harmonic oscillator coupled to a squeezed thermal bath, plotted as a function of the normalized squeezing parameter $R = |M| / \sqrt{N(N+1)}$. This dimensionless ratio quantifies the proximity of the bath state to the Heisenberg uncertainty bound. As $R$ increases, the anomalous correlations in the bath become stronger, and the steady-state of the oscillator becomes increasingly pure. In the limit $R \to 1$, corresponding to maximal squeezing allowed by quantum uncertainty, the system approaches a minimum-uncertainty squeezed state. However, for finite thermal occupation $N > 0$, full purity is never reached. In the specific case shown, with $N = 0.5$, the residual entropy reflects the mixed nature of the environment, even under ideal squeezing. This behavior is captured quantitatively through the R\'enyi-2 entropy, which provides a direct measure of purity via the determinant of the covariance matrix.

\subsection{Entropy Generation from Squeezing in Coupled Quantum Oscillators}

\begin{figure}
    \centering

\begin{tikzpicture}[scale=1, thick, every node/.style={font=\sffamily}]

  \draw[fill=gray!20] (-2.2,-0.7) rectangle (-1.2,0.7);
  \node at (-1.7, 0) {\small \( N_1, M_1 \)};
  
  \draw[fill=gray!20] (5.2,-0.7) rectangle (6.2,0.7);
  \node at (5.7, 0) {\small \( N_2, M_2 \)};
  
  \shade[ball color=blue!50!white] (0,0) circle (0.4);
  \node at (0,-.6) {\(\mathbf{a_1}\)};
  
  \shade[ball color=red!50!white] (4,0) circle (0.4);
 \node at (4,-.6) {\(\mathbf{a_2}\)};
  
  \draw[decorate,decoration={coil,aspect=0.5, segment length=6pt, amplitude=4pt}] (0.4,0) -- (3.6,0);
  \node[below] at (2, -0.2) {\( J \)};
  
  \draw[decorate,decoration={coil,aspect=0.5, segment length=6pt, amplitude=4pt}] (-1.2,0) -- (-0.4,0); 
 \draw[decorate,decoration={coil,aspect=0.5, segment length=6pt, amplitude=4pt}] (4.4,0) -- (5.2,0);   

\end{tikzpicture}
\caption{Schematic diagram of the two-mode system. Each harmonic oscillator (\( \mathbf{a}_1, \mathbf{a}_2 \)) is coupled to an independent squeezed thermal bath characterized by occupation \( N_i \) and squeezing parameter \( M_i \). The two oscillators interact via a coherent hopping term \( J \).}
    \label{fig:2HO}
\end{figure}
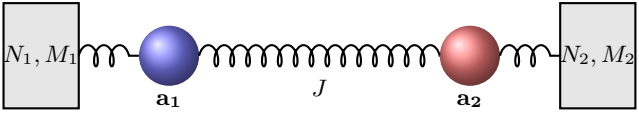

We now consider the possibility of using squeezed baths to perform thermodynamic work in open quantum systems. Specifically, we study the case of one or two harmonic oscillators, linearly coupled and each in contact with a local bosonic reservoir. Each bath is allowed to have its own thermal occupation number $N_i$ and squeezing amplitude $M_i$ (with $|M_i|^2 \leq N_i(N_i+1)$).
We denote the two system modes as $a_1$ and $a_2$, each with frequency $\omega_i$, and a coherent hopping term $J$ between them.  A sketch of this scenario is given 
in Fig.~\ref{fig:2HO} where oscillator 1 is connected to a reservoir on the left
while oscillator 2 is connected to the reservoir on the right. 
The total system Hamiltonian is:
\begin{equation}
H_S = \omega_1 a_1^\dagger a_1 + \omega_2 a_2^\dagger a_2 + J (a_1^\dagger a_2 + a_2^\dagger a_1).
\end{equation}

Each oscillator is coupled to its own local squeezed bath via the interaction:
\begin{equation}
H_{\text{int}} = \sum_{i=1}^2 \sum_k g_{i,k} (a_i b_{i,k}^\dagger + a_i^\dagger b_{i,k})
\end{equation}
where $b_{i,k}$ are the bath modes associated with oscillator $i$. 
Here, for the two-oscillator system, we define the covariance matrix,
\( \boldsymbol{\sigma} \) is given by the ensemble-averaged symmetrized outer product:
\begin{equation}
\boldsymbol{\sigma} = \langle \hat{\boldsymbol{R}} \hat{\boldsymbol{R}^\dagger}\rangle,
\end{equation}
where $\hat{\boldsymbol{R}}^\dagger =\{ \hat{a}_1^\dagger,\hat{a}_1,\hat{a}_2^\dagger,\hat{a}_2\}$. This yields a $4\times 4$
Hermitian matrix with populations $\langle \hat a_i^\dagger \hat a_i\rangle$, coherences
$\langle \hat a_1^\dagger \hat a_2\rangle$, and anomalous terms of the form $\langle \hat a_i \hat a_i\rangle$.
The resulting equations of motion take the usual linear form for the first moments
\begin{equation}
    \dv{\mathbf{R}}{t} = \mathbf{A}\cdot \mathbf{R}
\end{equation}
and a time-dependent Lyapunov equation
\begin{align}
    \dv{\boldsymbol{\sigma}}{t} = 
    \mathbf{A}\cdot\boldsymbol{\sigma} + \boldsymbol{\sigma}\cdot \mathbf{A}^\dagger +\mathbf{D},
\end{align}
for the covariances 
where $\mathbf{A}$ is the drift matrix derived from the linearized Heisenberg–Langevin equations and $
\mathbf{D}$ is the diffusion matrix determined by the bath noise correlations.  These matrices are given explicitly
by 
\begin{align}
    \mathbf{A} = 
    \left(
\begin{array}{cccc}
 -\frac{\gamma}{2}-i   \omega_1 & -\gamma M_1 & -i J & 0 \\
 -\gamma M_1 &    -\frac{\gamma}{2}+i    \omega_1 & 0 & i J \\
 -i J & 0 & -\frac{\gamma}{2}-i   \omega_2 & -\gamma M_2 \\
 0 & i J & -\gamma M_2    & -\frac{\gamma}{2}+i    \omega_2 \\
\end{array}
\right)
\label{eq:45}
\end{align}
and
\begin{align}
\mathbf{D} = 
\left(
\begin{array}{cccc}
 \gamma (N_1+2) & 2 \gamma   M_1 & 0 & 0 \\
 2 \gamma M_1 &  \gamma(N_1+2)   & 0 & 0 \\
 0 & 0 & \gamma(N_2+2)& 2 \gamma   M_2 \\
 0 & 0 & 2 \gamma M_2 &   \gamma(N_2 +2)\\
\end{array}
\right)
\end{align}
We shall use these quantities for the next analysis.

\paragraph{Exceptional Contours and Phase Structure}

Exceptional points in open quantum systems have recently been shown to assist squeezing and entanglement generation 
by controlling dissipation pathways \cite{Teixeira2023}. 
Moreover, a theoretical framework for Liouvillian exceptional points emerging from first principles has been proposed, 
clarifying their thermodynamic interpretation \cite{Khandelwal2024}.
In the broken-symmetry regime of the two-oscillator system, the eigenspectrum of the drift matrix $A$ reveals a distinct structure: the four eigenvalues naturally organize into two conjugate pairs, corresponding to symmetric and antisymmetric normal modes. This structure reflects the oscillator pair's underlying permutation symmetry. In the absence of squeezing, the dissipative evolution of each mode remains decoupled and exhibits no exceptional point (EP) structure—each eigenvalue remains non-degenerate and the system relaxes independently along its modal directions.

We now analyze the PT structure inherent in the two–oscillator system, using
the drift matrix $A$ already introduced in Eq.~\eqref{eq:45} 
and computing its eigenvalues. 
For a fixed dissipation rate $\gamma$, scanning the $(M_1,M_2)$ plane
reveals that the EPs form a family of rays emanating from the origin.  Each
ray corresponds to a specific squeezing phase (here encoded by the sign of
$M_1$ and $M_2$), and the full set of rays forms an ``exceptional–point
fan.''  Inside this fan both modes exhibit damped oscillatory dynamics
(PT–unbroken), while outside the fan one of the modes becomes purely
overdamped (PT–broken).  Increasing $\gamma$ broadens the fan, whereas
decreasing $\gamma$ collapses it to the coordinate axes.
This two–oscillator analysis provides a minimal and analytically tractable
framework for understanding the PT–breaking structure induced by squeezed
reservoirs.

A parity–time transformation acts as complex conjugation
($\mathcal{T}$) together with a mode–exchange operation ($\mathcal{P}$)
that interchanges $\hat{a}_i$ and $\hat{a}_i^\dagger$.  Since complex
conjugation reverses the sign of the anomalous correlators, the PT operation
maps
\begin{equation}
    M_i \longrightarrow -M_i.
\end{equation}
Consequently, the drift matrix $A$ can be PT–symmetric only if the squeezing
parameters satisfy
\begin{equation}
    M_1 M_2 < 0,
    \label{eq:PTcondition-twoosc}
\end{equation}
i.e., the two squeezed reservoirs must act on opposite quadratures.  Any
in–phase squeezing ($M_1 M_2 > 0$) explicitly breaks the PT symmetry of
$A$.  The origin $M_1=M_2=0$ is the only point that is invariant under the
PT operation without additional structure.

The condition~\eqref{eq:PTcondition-twoosc} partitions the $(M_1,M_2)$
plane into PT–broken and PT–unbroken regions.  In the first and third
quadrants ($M_1$ and $M_2$ of the same sign), the anomalous terms reinforce
each other and drive one Bogoliubov mode into a purely damped (non–oscillatory)
regime.  Thus, at least one eigenvalue of $A$ has zero imaginary part
throughout these quadrants, and the system is always in the PT–broken phase.
In contrast, the second and fourth quadrants ($M_1 M_2 < 0$) contain a
narrow wedge in which both normal modes retain nonzero oscillation
frequencies.  This region is PT–unbroken and emanates from the origin.  Its
boundaries correspond to exceptional points (EPs) at which two eigenvalues
of $A$ coalesce and the matrix becomes nondiagonalizable.  There is a second EP
involving the other normal mode of the system, which lines at unphysically 
large values of of $M_1$ and $M_2$. 

Crucially, this differs from the case of correlated thermal noise studied in our earlier work\cite{bittner2025clockworkquantumsymmetrynoise,bittner2025noiseinduceddecoherencefreezonesanyons,bittner2025statisticalcontrolrelaxationsynchronization,Bittner17072024,Bittner2025NoiseSyncOsc}, 
where noise correlations selectively protect a single normal mode, effectively locking its phase and making it immune to dissipation. In that case, the system exhibits global mode-locking and synchronization driven by the symmetric or antisymmetric structure of the environmental noise correlations.
Here, by contrast, local squeezing acts as a structured, information-rich reservoir rather than a source of correlated noise. Each bath breaks time-reversal symmetry locally via the squeezing phase, and the difference $M_1 - M_2$ provides a handle to tune how each normal mode decoheres and hybridizes with its environment. This leads to two distinct EPs, each corresponding to the coalescence of a different pair of eigenmodes—effectively a double bifurcation. In the broken-symmetry phase, both normal modes exhibit enhanced coherence and collective dynamics, yet remain independent: the system does not exhibit full mode-locking but rather partial synchronization across two dynamically distinct sectors.

\begin{figure}[t]
    \centering
    \includegraphics[width=0.75\linewidth]{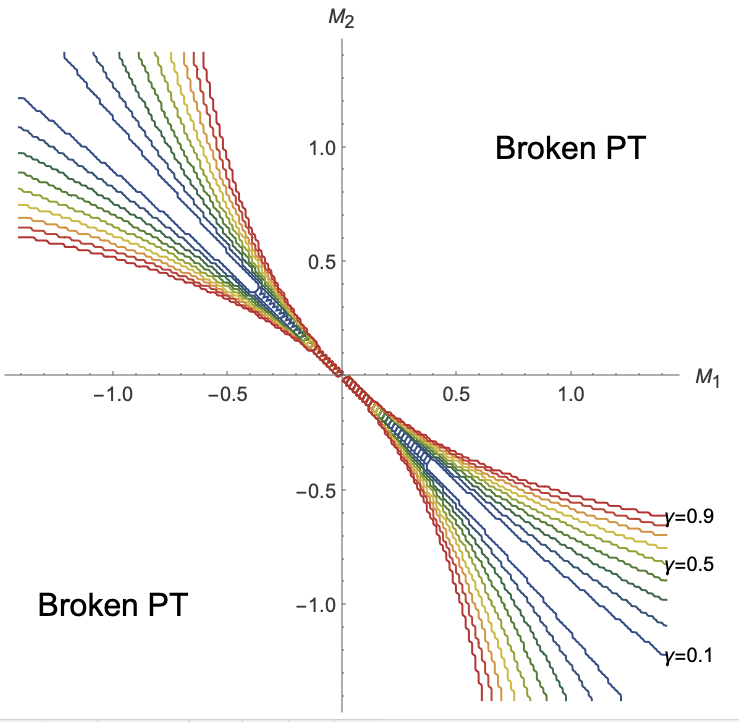}
    \caption{
        \textbf{PT–phase structure of the two–oscillator squeezed–bath model.}
        Shown are the exceptional–point (EP) contours in the $(M_{1},M_{2})$–plane
        for several values of the dissipation rate $\gamma$ (here
        $\gamma = 0.1,\,0.5,\,0.9$ from blue to red).
        Each curve marks the locus where the drift matrix
        $A$ (Eq.~(45)) becomes non-diagonalizable and two eigenvalues coalesce.
        Outside the colored ``fan’’ (first and third quadrants), the system lies in
        the \emph{broken} PT–symmetric phase, where one normal mode acquires
        purely real eigenvalues (overdamped, non-oscillatory dynamics).
        Inside the narrow region spanning the second and fourth quadrants,
        PT symmetry is \emph{unbroken}: both normal modes have complex-conjugate
        eigenvalues and oscillatory dynamics survive.
        The boundaries sharpen and contract as $\gamma$ decreases,
        showing that strong dissipation is required for EPs to occur within the
        physically allowed squeezing amplitudes.
    }
    \label{fig:PT_M1M2_fan}
\end{figure}


\paragraph{Role of the anomalous correlations}

In the absence of squeezing ($M_1 = M_2 = 0$), our  model reduces exactly to the configuration analyzed in our prior study of spontaneous synchronization in coupled quantum oscillators
\cite{Bittner2025NoiseSyncOsc} 
and one can obtain closed-form solutions for the steady-state Lyapunov equations for the symmetric case where 
$\omega_1 = \omega_2$ and $\gamma_1 = \gamma_2$. A key result from this is that 
population imbalance between local thermal reservoirs
$(\bar{n}_1 \ne \bar{n}_2)$ drives a unidirectional steady-state energy current between the two modes, governed by the cross-mode correlator
\begin{align}
   I^{(th)}_{1\to 2} =  \sigma_{23}+\sigma{32} =\frac{2J\gamma}{4J^2 + \gamma^2}(\bar{n}_2 - \bar{n}_1), 
\end{align}
which reflects transport mediated by the intermode coupling $J$ and balanced by dissipation into local baths at rate $\gamma$. This baseline captures purely thermal nonequilibrium flow in the absence of quantum correlations.  

For the case of a squeezed bath,  $M_{1,2}\ne 0 $,
we could not obtain a closed-form solution even for the 
symmetric case.  
However, we can obtain perturbative solutions
which should usually hold since the allowed values of $M$ are constrained by the
Heisenberg uncertainty principle. 
We introduce a dimensionless expansion parameter $\eta$ which is of order unity and write the drift and diffusion matrices as
\begin{align}
    A &= A^{(0)} + \eta A^{(1)} \qand  D &= D^{(0)} + \eta D^{(1)},
\end{align}
where $A^{(1)}$ and $D^{(1)}$ contain all the terms proportional to the squeezing parameters $M_1$ and $M_2$. The zero-order matrices $A^{(0)}$ and $D^{(0)}$ correspond to the case of thermal baths without squeezing ($M_i = 0$), for which the Lyapunov equation can be solved exactly.
This perturbative approach assumes “weak squeezing,” meaning that $|M_i| \ll \omega/\gamma$. As the system approaches the exceptional-point threshold $|M| \approx \omega/\gamma$, higher-order terms become significant and the linear expansion  ceases to be valid. In this strong-squeezing regime, numerical integration of the Lyapunov equation is required to capture the full non-perturbative dynamics.

As per Rayleigh-Schr\"odinger perturbation theory, 
we then expand the steady-state covariance matrix $\sigma$ in powers of $\eta$:
\begin{equation}
    \sigma = \sigma^{(0)} + \eta \sigma^{(1)} + \mathcal{O}(\eta^2),
\end{equation}
and insert this \textit{ansatz} into the Lyapunov equation.
Matching orders in $\eta$ yields a hierarchy of equations. At leading order ($\eta^0$), we recover the unperturbed solution:
\begin{equation}
    A^{(0)} \sigma^{(0)} + \sigma^{(0)} A^{(0)\dagger} = -D^{(0)}.
\end{equation}
\begin{widetext}
At first order in $\eta$, we obtain a linear matrix equation for $\sigma^{(1)}$:
\begin{equation}
    A^{(0)} \sigma^{(1)} + \sigma^{(1)} A^{(0)\dagger} = -\left( A^{(1)} \sigma^{(0)} + \sigma^{(0)} A^{(1)\dagger} + D^{(1)} \right).
    \label{eq:lyap-1st-order}
\end{equation}
Upon solving this, we note that the first-order correction to the covariance matrix due to asymmetric squeezing introduces anomalous cross-mode correlations: \( \sigma_{14}^{(1)} = \langle a_1^\dagger a_2^\dagger \rangle \) 
 given by 
 \begin{align}
 \sigma^{(1)}_{14} = \frac{2 i J \gamma \left(8 J^2 m (1+n) + \gamma \left[2 m(1+n) \gamma + \Delta m \Delta n (\gamma + i \omega)\right] \right)}{(4J^2 + \gamma^2)(4J^2 + (\gamma + 2i\omega)^2)}
 \end{align}
 normal cross-mode correlations:
\begin{align}
    \sigma^{(1)}_{23} = -\frac{2 i J \gamma \left(8 J^2 m (1+n) + \gamma \left[2 m(1+n) \gamma + \Delta m \Delta n (\gamma - i \omega)\right] \right)}{(4J^2 + \gamma^2)(4J^2 + (\gamma - 2i\omega)^2)}
\end{align} 
 where $n = (N_1+N_2)/2$ and $\Delta n = N_1-N_2$
 are the mean and differences of the effective occupations of the two baths, with similar expressions for $m$ and $\Delta m$.
 These terms, along with their Hermitian conjugates, populate the anti-diagonal of \( \sigma^{(1)} \), 
\begin{align}
\sigma^{(1)} =
\begin{pmatrix}
0 & 0 & 0 & \sigma_{14}^{(1)} \\
0 & 0 & \sigma_{23}^{(1)} & 0 \\
0 & \left( \sigma_{23}^{(1)} \right)^* & 0 & 0 \\
\left( \sigma_{14}^{(1)} \right)^* & 0 & 0 & 0
\end{pmatrix}.
\end{align}
This matrix is rank-2 and purely anti-diagonal. Its determinant becomes \( \left| \sigma_{14}^{(1)} \right|^4 \), leading to a compact Rényi-2 entropy correction:
\begin{align}
\Delta S_2^{(1)} \approx \frac{1}{2} \left| \sigma_{14}^{(1)} \right|^4.    
\end{align}
    
\end{widetext}

\paragraph{Renyi-2 Entropy Shift from Squeezing}

To quantify the contribution of squeezing to the steady-state entropy, we expand the determinant of the covariance matrix as $\sigma = \sigma_0 + \sigma_1 + \cdots$ and consider the Rényi-2 entropy correction:
\begin{equation}
\Delta S_2^{(\text{squeezing})} = \frac{1}{2} \log\left( \frac{\det \sigma}{\det \sigma_0} \right).
\end{equation}
Note that we choose to use the Rényi-2 entropy rather than the von Neumann entropy because it offers a closed-form expression for Gaussian states and is directly related to the state’s purity 
$P = \mathrm{Tr}[\rho^2]$. While Rényi-2 entropy may not capture all aspects of non-Gaussian correlations, it provides a robust quantitative measure of mixedness and coherence loss in linear bosonic systems.

Setting $A = \sigma_0^{-1} \cdot \sigma_1$ and using the identity
\[
\det(I + \varepsilon A) = 1 + \varepsilon \Tr(A) + \frac{\varepsilon^2}{2} \left[ \Tr(A)^2 - \Tr(A^2) \right] + \cdots,
\]
we find that the first-order contribution vanishes, since $\Tr(\sigma_0^{-1} \sigma_1) = 0$ by symmetry. Consequently, the leading-order entropy shift arises at second order:
\begin{equation}
\Delta S_2^{(\text{squeezing})} \approx -\frac{1}{4} \Tr\left[ (\sigma_0^{-1} \sigma_1)^2 \right].
\end{equation}
Taking $\Delta n = 0$ (i.e., no thermal bias), this reduces to a simple form:
\begin{equation}
\Delta S_2^{(\text{squeezing})} \propto \alpha_{n,m} + \beta_{n,m} \Delta m^2 + \cdots
\end{equation}
where $\alpha_{n,m} $ and $\beta_{n,m} $ are coefficients that depend on the average occupation and squeezing of the two opposing baths:
\begin{align}
\alpha_{n,m}  &= \frac{16 \gamma^2 m^2 (n+1)^2 \left( \gamma^2 + 4J^2 + 4\omega^2 \right)}{(2n+1)^2 \left[ \left( \gamma^2 + 4\omega^2 \right)^2 + 16J^4 + 8J^2 \left( \gamma^2 - 4\omega^2 \right) \right]} \\
\beta_{n,m}  &= \frac{4 \gamma^2 (n+1)^2}{(2n+1)^2 \left( \gamma^2 + 4\omega^2 \right)}.
\end{align}

This would indicate that entropy generation due to local squeezing is an intrinsically nonlinear effect. Importantly, the terms contributing to $\sigma_1$ originate from the anomalous correlations $\langle a_i a_j \rangle$ and their conjugates, which vanish in the absence of squeezing. Hence, while thermal gradients contribute linearly to entropy flow, coherence-induced entropy arises only at second order. This further highlights the distinction between heat transport and information-driven irreversibility in non-equilibrium quantum systems.
Here, “entropy generation” refers specifically to the increase of the system’s reduced-state entropy (e.g., Rényi-2 entropy) due to coherence exchange with structured environments. Unlike classical thermodynamic entropy production, this process does not require a net energy current; instead, it reflects the degradation of quantum information through non-unitary coupling to the squeezed reservoirs. In this sense, the entropy arises from the loss of accessible phase-space information rather than from dissipated heat.

This analysis also reveals a sharp distinction between entropy generation from incoherent thermal gradients and that induced by quantum coherence. While the former contributes linearly via $\Delta n$, the latter arises only at second order through $\sigma_1$, reflecting the fundamentally nonlinear role of anomalous correlations such as $\langle a_1^\dagger a_2^\dagger \rangle$. This result substantiates the notion that coherence is an independent thermodynamic resource—capable of producing entropy even in the absence of an energy gradient.

The quadratic structure of the entropy shift, in both $m$ and $\Delta m$, highlights the fundamentally nonlinear role of coherence in quantum thermodynamics. Unlike classical thermal entropy production—which arises from macroscopic energy gradients—the entropy generated here originates from microscopic phase-space correlations introduced by squeezed reservoirs. These correlations, encoded in the off-diagonal elements of the covariance matrix (e.g., $\sigma_{14}$), persist even when the baths have equal average occupation ($\Delta n = 0$). Thus, entropy production can occur even in the absence of heat flow.

This mechanism bears a striking resemblance to the modern reinterpretation of Maxwell’s demon: an entity that extracts work or induces directional flow not by injecting energy, but by manipulating information. In our system, the role of the demon is played by the squeezed baths, which encode phase-sensitive information in the form of nonzero anomalous second moments ($M_1$, $M_2$). The system's entropy increases not because energy is exchanged in the usual sense, but because coherence is transferred and degraded across the system. The breakdown of detailed balance occurs without a net energy current, reflecting an information-driven source of irreversibility.
The breakdown of detailed balance here does not stem from a net energy flow, but from an information-driven asymmetry introduced by phase-sensitive correlations in the reservoirs.

These results can be viewed through the lens of nonequilibrium fluctuation theorems for quantum maps, as developed by
Manzano, Horowitz, and Parrondo~\cite{Manzano2015}. Their framework formally connects entropy production in driven or
coherently biased systems—such as those coupled to squeezed baths—to generalized potentials that quantify deviations from
detailed balance.


In this context, the Rényi-2  entropy serves as a useful diagnostic: it captures not just thermal disorder, but the emergence of structured coherence and its nonlinear contribution to entropy. The result $\Delta S_2^{(\text{squeezing})} \sim |\sigma_{14}^{(1)}|^2$ suggests that entropy production is a signature of coherence flow—a measure of the system's inability to preserve or recover quantum information once entangled with asymmetric, phase-structured environments.
Our analysis underscores a broader conceptual shift in nonequilibrium physics: irreversibility and entropy are not solely tied to energy dissipation but can also stem from the loss of accessible quantum information. Local squeezing, therefore, becomes a probe of this deeper structure—where entropy generation is governed not by what is energetically exchanged, but by what is lost from the observer's informational vantage point.
The entropy production role of squeezing in non-equilibrium quantum engines has been recently revisited, 
demonstrating that phase-sensitive work extraction from squeezed baths parallels Maxwell-demon-like entropy control 
\cite{Arisoy2022}. 
This supports our interpretation of local squeezing as a coherent resource capable of transiently reversing entropy.

\section{Discussion: Squeezing, Information, and the Maxwell's Demon Analogy}

Our analysis reveals several physical effects that arise when local squeezed reservoirs couple to linearly interacting bosonic modes. Even without a thermal gradient, asymmetries in the squeezing parameters \( M_1 \) and \( M_2 \) generate steady-state currents and anomalous inter-mode correlations. This shows that squeezing introduces a hidden thermodynamic bias—one that drives directional transport of both energy and coherence.

A central technical result is the perturbative expansion of the steady-state covariance matrix \( \sigma = \sigma_0 + \sigma_1 + \cdots \). The anomalous contributions responsible for squeezing-induced entropy production appear first at second order. The resulting shift in the Rényi-2 entropy, which scales with \( |\sigma_{14}^{(1)}|^2 \), demonstrates that coherence-driven irreversibility is intrinsically nonlinear. While thermal gradients produce entropy at linear order, local quantum correlations do so only through higher-order processes. This contrast highlights a fundamental distinction between heat and coherence as thermodynamic resources.

We also identified exceptional points (EPs) in the eigenvalue spectrum of the drift matrix that signal transitions between distinct dynamical regimes. These EPs form contours in the space of squeezing parameters, delineating regions where eigenmode coalescence causes sharp changes in the system’s steady-state behavior. The resulting phase diagram exhibits a bifurcation structure: dynamical symmetries break when the asymmetry between baths exceeds a threshold determined by the system–bath coupling and oscillator parameters.

These findings invite comparison with the concept of a quantum Maxwell's Demon. The squeezing parameters \( M_i \) encode phase-sensitive information that reshapes bath statistics in ways inaccessible to classical thermometry. When these squeezing profiles differ, the system experiences a net flow of coherence and entropy—even when both baths have identical thermal occupations. The resulting anomalous correlations make this information-driven bias observable, showing how quantum coherence can replace classical energy gradients in producing directional transport. In this sense, squeezed reservoirs act like demons that use information to structure noise and generate work-like effects without net heat exchange.

Related frameworks in quantum stochastic thermodynamics—such as feedback-controlled demons~\cite{Esposito2009EPL} and information-driven work extraction~\cite{Strasberg2017PRX}—also interpret entropy production as a manifestation of informational asymmetry. Our analysis extends this perspective to continuously coupled squeezed reservoirs, where phase-sensitive correlations provide the informational bias that drives irreversible behavior.

These predictions lie within reach of current experimental platforms. Cavity optomechanical systems, superconducting circuits with engineered reservoirs, trapped ions illuminated by squeezed light, and hybrid light–matter systems in the microwave or THz domain all offer means of implementing controlled local squeezing. In superconducting circuit QED, typical parameters include oscillator frequencies \( \omega/2\pi \approx 5\,\mathrm{GHz} \), damping rates \( \gamma/2\pi \approx 1{-}10\,\mathrm{MHz} \), and achievable squeezing amplitudes \( r \approx 0.5{-}1.2 \) (corresponding to 4–10 dB of noise suppression). Within this regime, the predicted entropy shifts \( \Delta S_2 \sim 10^{-2} - 10^{-1} \) should be measurable via covariance-matrix tomography or noise spectral analysis.

Techniques for generating broadband squeezed vacuum are well established and can be integrated with tunable system–bath coupling and high-fidelity tomography to reconstruct the steady-state covariance matrix. Platforms that support reservoir engineering at the level of individual modes are especially promising, as they allow independent control of both the thermal and squeezed components of the bath. From a broader perspective, these results align with recent efforts to unify entropy generation, coherence, and feedback control in open quantum systems~\cite{Elouard2018,Manzano2015}. Together, these works suggest that squeezing and measurement-based feedback represent two complementary pathways toward information-powered thermodynamics.

In summary, local squeezing alters not only the bath's fluctuation spectrum but also the system's underlying thermodynamic structure. It drives coherence, entropy flow, and mode correlations, linking quantum information to nonequilibrium thermodynamics. These features lay the groundwork for future studies on entropy, irreversibility, and control in structured quantum environments.
\begin{acknowledgments}
The work at the University of Houston was supported by the National Science Foundation under CHE-2404788 and the Robert A. Welch Foundation (E-1337).

\end{acknowledgments}

\section*{Data Accessibility Statement.}
The author declares that the data supporting the findings of this study are available within the paper.

\section*{Authors contribution statement} 
ERB developed the concept, performed the theoretical analysis, and wrote the manuscript.

\section*{Conflicts of Interest}
The author has no conflicts of interest to declare.

\bibliographystyle{apsrev4-2}
\bibliography{squeeze_bath_refs}

\begin{thebibliography}{26}%
\makeatletter
\providecommand \@ifxundefined [1]{%
 \@ifx{#1\undefined}
}%
\providecommand \@ifnum [1]{%
 \ifnum #1\expandafter \@firstoftwo
 \else \expandafter \@secondoftwo
 \fi
}%
\providecommand \@ifx [1]{%
 \ifx #1\expandafter \@firstoftwo
 \else \expandafter \@secondoftwo
 \fi
}%
\providecommand \natexlab [1]{#1}%
\providecommand \enquote  [1]{``#1''}%
\providecommand \bibnamefont  [1]{#1}%
\providecommand \bibfnamefont [1]{#1}%
\providecommand \citenamefont [1]{#1}%
\providecommand \href@noop [0]{\@secondoftwo}%
\providecommand \href [0]{\begingroup \@sanitize@url \@href}%
\providecommand \@href[1]{\@@startlink{#1}\@@href}%
\providecommand \@@href[1]{\endgroup#1\@@endlink}%
\providecommand \@sanitize@url [0]{\catcode `\\12\catcode `\$12\catcode
  `\&12\catcode `\#12\catcode `\^12\catcode `\_12\catcode `\%12\relax}%
\providecommand \@@startlink[1]{}%
\providecommand \@@endlink[0]{}%
\providecommand \url  [0]{\begingroup\@sanitize@url \@url }%
\providecommand \@url [1]{\endgroup\@href {#1}{\urlprefix }}%
\providecommand \urlprefix  [0]{URL }%
\providecommand \Eprint [0]{\href }%
\providecommand \doibase [0]{https://doi.org/}%
\providecommand \selectlanguage [0]{\@gobble}%
\providecommand \bibinfo  [0]{\@secondoftwo}%
\providecommand \bibfield  [0]{\@secondoftwo}%
\providecommand \translation [1]{[#1]}%
\providecommand \BibitemOpen [0]{}%
\providecommand \bibitemStop [0]{}%
\providecommand \bibitemNoStop [0]{.\EOS\space}%
\providecommand \EOS [0]{\spacefactor3000\relax}%
\providecommand \BibitemShut  [1]{\csname bibitem#1\endcsname}%
\let\auto@bib@innerbib\@empty
\bibitem [{\citenamefont {Gardiner}\ and\ \citenamefont
  {Collett}(1985)}]{gardiner1985input}%
  \BibitemOpen
  \bibfield  {author} {\bibinfo {author} {\bibfnamefont {C.~W.}\ \bibnamefont
  {Gardiner}}\ and\ \bibinfo {author} {\bibfnamefont {M.~J.}\ \bibnamefont
  {Collett}},\ }\href {https://doi.org/10.1103/PhysRevA.31.3761} {\bibfield
  {journal} {\bibinfo  {journal} {Physical Review A}\ }\textbf {\bibinfo
  {volume} {31}},\ \bibinfo {pages} {3761} (\bibinfo {year}
  {1985})}\BibitemShut {NoStop}%
\bibitem [{\citenamefont {Caves}(1981)}]{caves1981quantum}%
  \BibitemOpen
  \bibfield  {author} {\bibinfo {author} {\bibfnamefont {C.~M.}\ \bibnamefont
  {Caves}},\ }\href {https://doi.org/10.1103/PhysRevD.23.1693} {\bibfield
  {journal} {\bibinfo  {journal} {Physical Review D}\ }\textbf {\bibinfo
  {volume} {23}},\ \bibinfo {pages} {1693} (\bibinfo {year}
  {1981})}\BibitemShut {NoStop}%
\bibitem [{\citenamefont {Kronwald}\ \emph {et~al.}(2013)\citenamefont
  {Kronwald}, \citenamefont {Marquardt},\ and\ \citenamefont
  {Clerk}}]{kronwald2013arbitrarily}%
  \BibitemOpen
  \bibfield  {author} {\bibinfo {author} {\bibfnamefont {A.}~\bibnamefont
  {Kronwald}}, \bibinfo {author} {\bibfnamefont {F.}~\bibnamefont
  {Marquardt}},\ and\ \bibinfo {author} {\bibfnamefont {A.~A.}\ \bibnamefont
  {Clerk}},\ }\href {https://doi.org/10.1103/PhysRevA.88.063833} {\bibfield
  {journal} {\bibinfo  {journal} {Physical Review A}\ }\textbf {\bibinfo
  {volume} {88}},\ \bibinfo {pages} {063833} (\bibinfo {year}
  {2013})}\BibitemShut {NoStop}%
\bibitem [{\citenamefont {Metelmann}\ and\ \citenamefont
  {Clerk}(2015)}]{metelmann2015nonreciprocal}%
  \BibitemOpen
  \bibfield  {author} {\bibinfo {author} {\bibfnamefont {A.}~\bibnamefont
  {Metelmann}}\ and\ \bibinfo {author} {\bibfnamefont {A.~A.}\ \bibnamefont
  {Clerk}},\ }\href {https://doi.org/10.1103/PhysRevX.5.021025} {\bibfield
  {journal} {\bibinfo  {journal} {Physical Review X}\ }\textbf {\bibinfo
  {volume} {5}},\ \bibinfo {pages} {021025} (\bibinfo {year}
  {2015})}\BibitemShut {NoStop}%
\bibitem [{\citenamefont {Arısoy}\ \emph {et~al.}(2022)\citenamefont
  {Arısoy}, \citenamefont {Hsiang},\ and\ \citenamefont {Hu}}]{Arisoy2022}%
  \BibitemOpen
  \bibfield  {author} {\bibinfo {author} {\bibfnamefont {O.}~\bibnamefont
  {Arısoy}}, \bibinfo {author} {\bibfnamefont {J.~T.}\ \bibnamefont
  {Hsiang}},\ and\ \bibinfo {author} {\bibfnamefont {B.~L.}\ \bibnamefont
  {Hu}},\ }\href {https://doi.org/10.1103/PhysRevE.105.014108} {\bibfield
  {journal} {\bibinfo  {journal} {Physical Review E}\ }\textbf {\bibinfo
  {volume} {105}},\ \bibinfo {pages} {014108} (\bibinfo {year}
  {2022})}\BibitemShut {NoStop}%
\bibitem [{\citenamefont {Mallet}\ \emph {et~al.}(2011)\citenamefont {Mallet},
  \citenamefont {Castellanos-Beltran}, \citenamefont {Ku}, \citenamefont
  {Glancy}, \citenamefont {Knill}, \citenamefont {Irwin}, \citenamefont
  {Hilton}, \citenamefont {Vale},\ and\ \citenamefont
  {Lehnert}}]{mallet2011quantum}%
  \BibitemOpen
  \bibfield  {author} {\bibinfo {author} {\bibfnamefont {F.}~\bibnamefont
  {Mallet}}, \bibinfo {author} {\bibfnamefont {M.~A.}\ \bibnamefont
  {Castellanos-Beltran}}, \bibinfo {author} {\bibfnamefont {H.}~\bibnamefont
  {Ku}}, \bibinfo {author} {\bibfnamefont {S.}~\bibnamefont {Glancy}}, \bibinfo
  {author} {\bibfnamefont {E.}~\bibnamefont {Knill}}, \bibinfo {author}
  {\bibfnamefont {K.~D.}\ \bibnamefont {Irwin}}, \bibinfo {author}
  {\bibfnamefont {G.~C.}\ \bibnamefont {Hilton}}, \bibinfo {author}
  {\bibfnamefont {L.~R.}\ \bibnamefont {Vale}},\ and\ \bibinfo {author}
  {\bibfnamefont {K.~W.}\ \bibnamefont {Lehnert}},\ }\href
  {https://doi.org/10.1103/PhysRevLett.106.220502} {\bibfield  {journal}
  {\bibinfo  {journal} {Physical Review Letters}\ }\textbf {\bibinfo {volume}
  {106}},\ \bibinfo {pages} {220502} (\bibinfo {year} {2011})}\BibitemShut
  {NoStop}%
\bibitem [{\citenamefont {Flurin}\ \emph {et~al.}(2012)\citenamefont {Flurin},
  \citenamefont {Roch}, \citenamefont {Mallet}, \citenamefont {Devoret},\ and\
  \citenamefont {Huard}}]{flurin2012generating}%
  \BibitemOpen
  \bibfield  {author} {\bibinfo {author} {\bibfnamefont {E.}~\bibnamefont
  {Flurin}}, \bibinfo {author} {\bibfnamefont {N.}~\bibnamefont {Roch}},
  \bibinfo {author} {\bibfnamefont {F.}~\bibnamefont {Mallet}}, \bibinfo
  {author} {\bibfnamefont {M.~H.}\ \bibnamefont {Devoret}},\ and\ \bibinfo
  {author} {\bibfnamefont {B.}~\bibnamefont {Huard}},\ }\href
  {https://doi.org/10.1103/PhysRevLett.109.183901} {\bibfield  {journal}
  {\bibinfo  {journal} {Physical Review Letters}\ }\textbf {\bibinfo {volume}
  {109}},\ \bibinfo {pages} {183901} (\bibinfo {year} {2012})}\BibitemShut
  {NoStop}%
\bibitem [{\citenamefont {Gardiner}\ and\ \citenamefont
  {Zoller}(2004)}]{Gardiner2004}%
  \BibitemOpen
  \bibfield  {author} {\bibinfo {author} {\bibfnamefont {C.~W.}\ \bibnamefont
  {Gardiner}}\ and\ \bibinfo {author} {\bibfnamefont {P.}~\bibnamefont
  {Zoller}},\ }\href {https://doi.org/10.1007/978-3-540-30183-4} {\emph
  {\bibinfo {title} {Quantum Noise: A Handbook of Markovian and Non-Markovian
  Quantum Stochastic Methods with Applications to Quantum Optics}}},\ \bibinfo
  {edition} {3rd}\ ed.,\ Springer Series in Synergetics\ (\bibinfo  {publisher}
  {Springer},\ \bibinfo {year} {2004})\BibitemShut {NoStop}%
\bibitem [{\citenamefont {Zippilli}\ and\ \citenamefont
  {Illuminati}(2014)}]{Zippilli2014}%
  \BibitemOpen
  \bibfield  {author} {\bibinfo {author} {\bibfnamefont {S.}~\bibnamefont
  {Zippilli}}\ and\ \bibinfo {author} {\bibfnamefont {F.}~\bibnamefont
  {Illuminati}},\ }\href {https://doi.org/10.1103/PhysRevA.89.033803}
  {\bibfield  {journal} {\bibinfo  {journal} {Physical Review A}\ }\textbf
  {\bibinfo {volume} {89}},\ \bibinfo {pages} {033803} (\bibinfo {year}
  {2014})}\BibitemShut {NoStop}%
\bibitem [{\citenamefont {Murch}\ \emph {et~al.}(2013)\citenamefont {Murch},
  \citenamefont {Vijay}, \citenamefont {Macklin},\ and\ \citenamefont
  {Siddiqi}}]{Murch2013}%
  \BibitemOpen
  \bibfield  {author} {\bibinfo {author} {\bibfnamefont {K.~W.}\ \bibnamefont
  {Murch}}, \bibinfo {author} {\bibfnamefont {R.}~\bibnamefont {Vijay}},
  \bibinfo {author} {\bibfnamefont {C.}~\bibnamefont {Macklin}},\ and\ \bibinfo
  {author} {\bibfnamefont {I.}~\bibnamefont {Siddiqi}},\ }\href
  {https://doi.org/10.1038/nature12264} {\bibfield  {journal} {\bibinfo
  {journal} {Nature}\ }\textbf {\bibinfo {volume} {499}},\ \bibinfo {pages}
  {62} (\bibinfo {year} {2013})}\BibitemShut {NoStop}%
\bibitem [{\citenamefont {Klaers}\ \emph {et~al.}(2017)\citenamefont {Klaers},
  \citenamefont {Faelt}, \citenamefont {Imamoglu},\ and\ \citenamefont
  {Togan}}]{Klaers2017}%
  \BibitemOpen
  \bibfield  {author} {\bibinfo {author} {\bibfnamefont {J.}~\bibnamefont
  {Klaers}}, \bibinfo {author} {\bibfnamefont {S.}~\bibnamefont {Faelt}},
  \bibinfo {author} {\bibfnamefont {A.}~\bibnamefont {Imamoglu}},\ and\
  \bibinfo {author} {\bibfnamefont {E.}~\bibnamefont {Togan}},\ }\href
  {https://doi.org/10.1103/PhysRevX.7.031044} {\bibfield  {journal} {\bibinfo
  {journal} {Physical Review X}\ }\textbf {\bibinfo {volume} {7}},\ \bibinfo
  {pages} {031044} (\bibinfo {year} {2017})}\BibitemShut {NoStop}%
\bibitem [{\citenamefont {Alicki}\ and\ \citenamefont
  {Gelbwaser-Klimovsky}(2015)}]{Alicki2015}%
  \BibitemOpen
  \bibfield  {author} {\bibinfo {author} {\bibfnamefont {R.}~\bibnamefont
  {Alicki}}\ and\ \bibinfo {author} {\bibfnamefont {D.}~\bibnamefont
  {Gelbwaser-Klimovsky}},\ }\href
  {https://doi.org/10.1088/1367-2630/17/11/115012} {\bibfield  {journal}
  {\bibinfo  {journal} {New Journal of Physics}\ }\textbf {\bibinfo {volume}
  {17}},\ \bibinfo {pages} {115012} (\bibinfo {year} {2015})}\BibitemShut
  {NoStop}%
\bibitem [{\citenamefont {Heiss}(2012)}]{Heiss2012}%
  \BibitemOpen
  \bibfield  {author} {\bibinfo {author} {\bibfnamefont {W.~D.}\ \bibnamefont
  {Heiss}},\ }\href {https://doi.org/10.1088/1751-8113/45/44/444016} {\bibfield
   {journal} {\bibinfo  {journal} {Journal of Physics A: Mathematical and
  Theoretical}\ }\textbf {\bibinfo {volume} {45}},\ \bibinfo {pages} {444016}
  (\bibinfo {year} {2012})}\BibitemShut {NoStop}%
\bibitem [{\citenamefont {Minganti}\ \emph {et~al.}(2019)\citenamefont
  {Minganti}, \citenamefont {Miranowicz}, \citenamefont {Chhajlany},\ and\
  \citenamefont {Nori}}]{Minganti2019}%
  \BibitemOpen
  \bibfield  {author} {\bibinfo {author} {\bibfnamefont {F.}~\bibnamefont
  {Minganti}}, \bibinfo {author} {\bibfnamefont {A.}~\bibnamefont
  {Miranowicz}}, \bibinfo {author} {\bibfnamefont {R.~W.}\ \bibnamefont
  {Chhajlany}},\ and\ \bibinfo {author} {\bibfnamefont {F.}~\bibnamefont
  {Nori}},\ }\href {https://doi.org/10.1103/PhysRevA.100.062131} {\bibfield
  {journal} {\bibinfo  {journal} {Physical Review A}\ }\textbf {\bibinfo
  {volume} {100}},\ \bibinfo {pages} {062131} (\bibinfo {year}
  {2019})}\BibitemShut {NoStop}%
\bibitem [{\citenamefont {Ablimit}\ \emph {et~al.}(2023)\citenamefont
  {Ablimit}, \citenamefont {Ren}, \citenamefont {He}, \citenamefont {Xie},\
  and\ \citenamefont {Wang}}]{Ablimit2023}%
  \BibitemOpen
  \bibfield  {author} {\bibinfo {author} {\bibfnamefont {A.}~\bibnamefont
  {Ablimit}}, \bibinfo {author} {\bibfnamefont {F.~H.}\ \bibnamefont {Ren}},
  \bibinfo {author} {\bibfnamefont {R.~H.}\ \bibnamefont {He}}, \bibinfo
  {author} {\bibfnamefont {Y.~Y.}\ \bibnamefont {Xie}},\ and\ \bibinfo {author}
  {\bibfnamefont {Z.~M.}\ \bibnamefont {Wang}},\ }\href
  {https://www.sciencedirect.com/science/article/pii/S0378437123008063}
  {\bibfield  {journal} {\bibinfo  {journal} {Physica A: Statistical Mechanics
  and its Applications}\ } (\bibinfo {year} {2023})}\BibitemShut {NoStop}%
\bibitem [{\citenamefont {Teixeira}\ \emph {et~al.}(2023)\citenamefont
  {Teixeira}, \citenamefont {Vadimov}, \citenamefont {M{\"o}rstedt},\ and\
  \citenamefont {Kundu}}]{Teixeira2023}%
  \BibitemOpen
  \bibfield  {author} {\bibinfo {author} {\bibfnamefont {W.~S.}\ \bibnamefont
  {Teixeira}}, \bibinfo {author} {\bibfnamefont {V.}~\bibnamefont {Vadimov}},
  \bibinfo {author} {\bibfnamefont {T.}~\bibnamefont {M{\"o}rstedt}},\ and\
  \bibinfo {author} {\bibfnamefont {S.}~\bibnamefont {Kundu}},\ }\href
  {https://doi.org/10.1103/PhysRevResearch.5.033119} {\bibfield  {journal}
  {\bibinfo  {journal} {Physical Review Research}\ }\textbf {\bibinfo {volume}
  {5}},\ \bibinfo {pages} {033119} (\bibinfo {year} {2023})}\BibitemShut
  {NoStop}%
\bibitem [{\citenamefont {Khandelwal}\ and\ \citenamefont
  {Blasi}(2024)}]{Khandelwal2024}%
  \BibitemOpen
  \bibfield  {author} {\bibinfo {author} {\bibfnamefont {S.}~\bibnamefont
  {Khandelwal}}\ and\ \bibinfo {author} {\bibfnamefont {G.}~\bibnamefont
  {Blasi}},\ }\href {https://arxiv.org/abs/2409.08100} {\bibfield  {journal}
  {\bibinfo  {journal} {arXiv preprint}\ } (\bibinfo {year} {2024})},\ \Eprint
  {https://arxiv.org/abs/2409.08100} {2409.08100} \BibitemShut {NoStop}%
\bibitem [{\citenamefont {Bittner}\ and\ \citenamefont
  {Tyagi}(2025{\natexlab{a}})}]{bittner2025clockworkquantumsymmetrynoise}%
  \BibitemOpen
  \bibfield  {author} {\bibinfo {author} {\bibfnamefont {E.~R.}\ \bibnamefont
  {Bittner}}\ and\ \bibinfo {author} {\bibfnamefont {B.}~\bibnamefont
  {Tyagi}},\ }\href {https://arxiv.org/abs/2507.19348} {\  (\bibinfo {year}
  {2025}{\natexlab{a}})},\ \Eprint {https://arxiv.org/abs/2507.19348}
  {arXiv:2507.19348 [quant-ph]} \BibitemShut {NoStop}%
\bibitem [{\citenamefont
  {Bittner}(2025)}]{bittner2025noiseinduceddecoherencefreezonesanyons}%
  \BibitemOpen
  \bibfield  {author} {\bibinfo {author} {\bibfnamefont {E.~R.}\ \bibnamefont
  {Bittner}},\ }\href {https://arxiv.org/abs/2510.06094} {\  (\bibinfo {year}
  {2025})},\ \Eprint {https://arxiv.org/abs/2510.06094} {arXiv:2510.06094
  [quant-ph]} \BibitemShut {NoStop}%
\bibitem [{\citenamefont {Bittner}\ and\ \citenamefont
  {Tyagi}(2025{\natexlab{b}})}]{bittner2025statisticalcontrolrelaxationsynchronization}%
  \BibitemOpen
  \bibfield  {author} {\bibinfo {author} {\bibfnamefont {E.~R.}\ \bibnamefont
  {Bittner}}\ and\ \bibinfo {author} {\bibfnamefont {B.}~\bibnamefont
  {Tyagi}},\ }\href {https://arxiv.org/abs/2504.02173} {\bibfield  {journal}
  {\bibinfo  {journal} {Scientific Reports}\ } (\bibinfo {year} {in press
  2025}{\natexlab{b}})},\ \Eprint {https://arxiv.org/abs/2504.02173}
  {arXiv:2504.02173 [quant-ph]} \BibitemShut {NoStop}%
\bibitem [{\citenamefont {Bittner}\ \emph {et~al.}(2024)\citenamefont
  {Bittner}, \citenamefont {Li}, \citenamefont {Shah}, \citenamefont
  {Silva-Acu{\~n}a},\ and\ \citenamefont {Piryatinski}}]{Bittner17072024}%
  \BibitemOpen
  \bibfield  {author} {\bibinfo {author} {\bibfnamefont {E.~R.}\ \bibnamefont
  {Bittner}}, \bibinfo {author} {\bibfnamefont {H.}~\bibnamefont {Li}},
  \bibinfo {author} {\bibfnamefont {S.~A.}\ \bibnamefont {Shah}}, \bibinfo
  {author} {\bibfnamefont {C.}~\bibnamefont {Silva-Acu{\~n}a}},\ and\ \bibinfo
  {author} {\bibfnamefont {A.}~\bibnamefont {Piryatinski}},\ }\href
  {https://doi.org/10.1080/14786435.2024.2341011} {\bibfield  {journal}
  {\bibinfo  {journal} {Philosophical Magazine}\ }\textbf {\bibinfo {volume}
  {104}},\ \bibinfo {pages} {630} (\bibinfo {year} {2024})}\BibitemShut
  {NoStop}%
\bibitem [{\citenamefont {Bittner}\ and\ \citenamefont
  {Tyagi}(2025{\natexlab{c}})}]{Bittner2025NoiseSyncOsc}%
  \BibitemOpen
  \bibfield  {author} {\bibinfo {author} {\bibfnamefont {E.~R.}\ \bibnamefont
  {Bittner}}\ and\ \bibinfo {author} {\bibfnamefont {B.}~\bibnamefont
  {Tyagi}},\ }\href {https://doi.org/10.1063/5.0246275} {\bibfield  {journal}
  {\bibinfo  {journal} {The Journal of Chemical Physics}\ }\textbf {\bibinfo
  {volume} {162}},\ \bibinfo {pages} {104116} (\bibinfo {year}
  {2025}{\natexlab{c}})}\BibitemShut {NoStop}%
\bibitem [{\citenamefont {Manzano}\ \emph {et~al.}(2015)\citenamefont
  {Manzano}, \citenamefont {Horowitz},\ and\ \citenamefont
  {Parrondo}}]{Manzano2015}%
  \BibitemOpen
  \bibfield  {author} {\bibinfo {author} {\bibfnamefont {G.}~\bibnamefont
  {Manzano}}, \bibinfo {author} {\bibfnamefont {J.~M.}\ \bibnamefont
  {Horowitz}},\ and\ \bibinfo {author} {\bibfnamefont {J.~M.~R.}\ \bibnamefont
  {Parrondo}},\ }\href {https://doi.org/10.1088/1367-2630/17/9/093029}
  {\bibfield  {journal} {\bibinfo  {journal} {New Journal of Physics}\ }\textbf
  {\bibinfo {volume} {17}},\ \bibinfo {pages} {093029} (\bibinfo {year}
  {2015})}\BibitemShut {NoStop}%
\bibitem [{\citenamefont {Esposito}\ and\ \citenamefont {Van~den
  Broeck}(2009)}]{Esposito2009EPL}%
  \BibitemOpen
  \bibfield  {author} {\bibinfo {author} {\bibfnamefont {M.}~\bibnamefont
  {Esposito}}\ and\ \bibinfo {author} {\bibfnamefont {C.}~\bibnamefont {Van~den
  Broeck}},\ }\href {https://doi.org/10.1209/0295-5075/85/60010} {\bibfield
  {journal} {\bibinfo  {journal} {Europhysics Letters}\ }\textbf {\bibinfo
  {volume} {85}},\ \bibinfo {pages} {60010} (\bibinfo {year}
  {2009})}\BibitemShut {NoStop}%
\bibitem [{\citenamefont {Strasberg}\ \emph {et~al.}(2017)\citenamefont
  {Strasberg}, \citenamefont {Schaller}, \citenamefont {Brandes},\ and\
  \citenamefont {Esposito}}]{Strasberg2017PRX}%
  \BibitemOpen
  \bibfield  {author} {\bibinfo {author} {\bibfnamefont {P.}~\bibnamefont
  {Strasberg}}, \bibinfo {author} {\bibfnamefont {G.}~\bibnamefont {Schaller}},
  \bibinfo {author} {\bibfnamefont {T.}~\bibnamefont {Brandes}},\ and\ \bibinfo
  {author} {\bibfnamefont {M.}~\bibnamefont {Esposito}},\ }\href
  {https://doi.org/10.1103/PhysRevX.7.021003} {\bibfield  {journal} {\bibinfo
  {journal} {Physical Review X}\ }\textbf {\bibinfo {volume} {7}},\ \bibinfo
  {pages} {021003} (\bibinfo {year} {2017})}\BibitemShut {NoStop}%
\bibitem [{\citenamefont {Elouard}\ and\ \citenamefont
  {Jordan}(2018)}]{Elouard2018}%
  \BibitemOpen
  \bibfield  {author} {\bibinfo {author} {\bibfnamefont {C.}~\bibnamefont
  {Elouard}}\ and\ \bibinfo {author} {\bibfnamefont {A.~N.}\ \bibnamefont
  {Jordan}},\ }\href {https://doi.org/10.1103/PhysRevLett.120.260601}
  {\bibfield  {journal} {\bibinfo  {journal} {Physical Review Letters}\
  }\textbf {\bibinfo {volume} {120}},\ \bibinfo {pages} {260601} (\bibinfo
  {year} {2018})}\BibitemShut {NoStop}%
\end{thebibliography}%

\newpage
\appendix

\end{document}